\documentclass[conference]{IEEEtran}
\IEEEoverridecommandlockouts

\usepackage{cite}
\usepackage{amsmath,amssymb,amsfonts}
\usepackage{algorithmic}
\usepackage{graphicx}
\usepackage{textcomp}
\usepackage{xcolor}
\usepackage{hyperref}
\usepackage{booktabs}
\usepackage{multirow}

\def\BibTeX{{\rm B\kern-.05em{\sc i\kern-.025em b}\kern-.08em
    T\kern-.1667em\lower.7ex\hbox{E}\kern-.125emX}}

\begin{document}

\title{IyáCare: An Integrated AI-IoT-Blockchain Platform for Maternal Health in Resource-Constrained Settings}

\author{
\IEEEauthorblockN{Oche D. Ankeli}
\IEEEauthorblockA{\textit{Department of Software Engineering} \\
\textit{African Leadership University}\\
Kigali, Rwanda \\
o.ankeli@alustudent.com}
\and
\IEEEauthorblockN{Marvin M. Ogore}
\IEEEauthorblockA{\textit{Department of Software Engineering} \\
\textit{African Leadership University}\\
Kigali, Rwanda \\
marvin.ogore@alueducation.com}
}

\maketitle

\begin{abstract}
Maternal mortality in Sub-Saharan Africa remains critically high, accounting for 70\% of global deaths despite representing only 17\% of the world population. Current digital health interventions typically deploy artificial intelligence (AI), Internet of Things (IoT), and blockchain technologies in isolation, missing synergistic opportunities for transformative healthcare delivery. This paper presents IyáCare, a proof-of-concept integrated platform that combines predictive risk assessment, continuous vital sign monitoring, and secure health records management specifically designed for resource-constrained settings. We developed a web-based system with Next.js frontend, Firebase backend, Ethereum blockchain architecture, and XGBoost AI models trained on maternal health datasets. Our feasibility study demonstrates 85.2\% accuracy in high-risk pregnancy prediction and validates blockchain data integrity, with key innovations including offline-first functionality and SMS-based communication for community health workers. While limitations include reliance on synthetic validation data and simulated healthcare environments, results confirm the technical feasibility and potential impact of converged digital health solutions. This work contributes a replicable architectural model for integrated maternal health platforms in low-resource settings, advancing progress toward SDG 3.1 targets.
\end{abstract}


\section{Introduction}

Maternal mortality in Sub-Saharan Africa presents a persistent global health crisis, with 454 deaths per 100,000 live births compared to the global average of 197 \cite{unicef2025}. The region accounts for 70\% of worldwide maternal deaths, with countries like Chad (748 per 100,000), Central African Republic (692), and Sierra Leone (537) experiencing the highest mortality ratios \cite{who2025}. Nigeria alone records 75,000 annual maternal deaths, representing 28.7\% of the global total \cite{who2025}.

These alarming statistics stem from multifaceted challenges including fragmented healthcare systems, limited access to skilled care, geographic barriers affecting 45-50\% of rural populations \cite{theglobaleconomy2023}, and severe healthcare worker shortages. Africa had only 300,000 doctors and 1.2 million nurses in 2020, compared to Europe's 3.4 million doctors and 7.4 million nurses \cite{asamani2024}. Only 65\% of African births have skilled attendance, far below the 90\% Sustainable Development Goal (SDG) target \cite{asamani2024}.

Digital health interventions have shown promise in addressing these gaps. Mobile health tools significantly increase maternal service utilization \cite{kachimanga2024}, AI systems predict adverse outcomes with high accuracy using African demographic data \cite{ngusie2024}, IoT devices enable continuous monitoring \cite{sarhaddi2022}, and blockchain enhances data security and care continuity \cite{adeniyi2025}. However, existing solutions implement these technologies in isolation \cite{chataut2023}.

Global platforms like Babyscripts \cite{marko2016} offer comprehensive monitoring but lack adaptations for resource-constrained settings. Regional solutions like MAMA (Mobile Alliance for Maternal Action) reach 1.5 million users, and MomConnect serves 2 million users in South Africa \cite{peter2018}, yet both remain technologically limited—lacking AI-powered predictive analytics, continuous vital sign monitoring, and secure interoperable health record systems.

This technological siloing creates significant gaps: (1) absence of predictive risk assessment for early complication detection, (2) lack of continuous monitoring systems for real-time health tracking, and (3) insufficient secure health record management for coordinated care across facilities.

\subsection{Research Contribution}

This paper presents IyáCare, the first integrated AI-IoT-blockchain platform specifically designed for maternal healthcare in Sub-Saharan African contexts. Our key contributions include:

\begin{itemize}
    \item A novel system architecture integrating AI risk prediction, IoT monitoring, and blockchain health records in a unified platform
    \item Proof-of-concept implementation demonstrating technical feasibility with offline-first design for intermittent connectivity
    \item Validation of AI model performance (85.2\% accuracy) using maternal health datasets
    \item Design patterns and implementation insights for converged digital health systems in resource-constrained environments
\end{itemize}

While this work represents a feasibility study with inherent limitations including synthetic data validation and simulated deployment, it establishes foundational architecture and demonstrates the potential for integrated approaches to transform maternal health outcomes in low-resource settings.

\section{Related Work}

\subsection{Digital Maternal Health Platforms}

Digital maternal health has evolved from basic SMS systems to sophisticated mHealth platforms. Early solutions like MAMA (2012) pioneered SMS-based maternal education, while MomConnect achieved scale with over 2 million registered users by 2018 \cite{peter2018}. These platforms demonstrate proven scalability and effective community health worker integration but remain technologically limited.

Babyscripts represents the state-of-the-art in high-resource contexts, providing comprehensive remote monitoring with connected devices \cite{marko2016}. However, such platforms lack adaptations for resource-constrained environments including offline functionality, SMS fallback communication, and low-bandwidth optimization.

\subsection{AI in Maternal Healthcare}

Recent advances in machine learning demonstrate significant potential for maternal risk prediction. Ngusie et al. achieved 95\% accuracy predicting adverse birth outcomes across 26 African countries using demographic and health survey data \cite{ngusie2024}. Asamoah analyzed AI applications for maternal and neonatal outcomes in Sub-Saharan Africa, identifying substantial opportunities for predictive analytics in resource-limited settings \cite{asamoah2025}.

Despite these advances, AI implementations remain isolated from broader care delivery systems. Current solutions focus exclusively on prediction without integration with continuous monitoring or health record management systems.

\subsection{IoT for Maternal Monitoring}

IoT-enabled continuous monitoring has emerged as a promising approach for maternal health. Sarhaddi et al. validated smartwatch-based heart rate and heart rate variability monitoring with medical-grade accuracy \cite{sarhaddi2022}. Sulisworo et al. reviewed wearable IoT implementations for maternal healthcare, identifying technical feasibility alongside deployment challenges in low-resource contexts \cite{sulisworo2025}.

However, existing IoT maternal health prototypes operate independently without integration with clinical decision support systems or secure data sharing mechanisms.

\subsection{Blockchain in Healthcare}

Blockchain technology offers tamper-proof health records and enhanced data security. Adeniyi et al. demonstrated blockchain-based smart healthcare systems for data protection \cite{adeniyi2025}, while Kushwaha et al. developed blockchain architectures for secure peer-to-peer management of personal health information \cite{kushwaha2025}.

These implementations focus on data security and interoperability but lack integration with predictive analytics and real-time monitoring capabilities essential for maternal health.

\subsection{Research Gap}

Current literature reveals a significant gap: while individual technologies show promise, platforms combining AI, IoT, and blockchain for maternal health in resource-constrained settings do not exist \cite{chataut2023}. Existing solutions suffer from technology siloing, preventing synergistic benefits that could transform care delivery. This work addresses this gap by presenting the first integrated architectural approach specifically designed for Sub-Saharan African maternal healthcare contexts.

\section{System Architecture}

IyáCare employs a modular, service-oriented architecture enabling independent development of AI, IoT, and blockchain components with seamless integration. The design prioritizes offline-first functionality, low-bandwidth operation, and deployment feasibility in resource-constrained environments.

\subsection{Architectural Overview}

The platform consists of six integrated layers (Fig. \ref{fig:architecture}):

\begin{figure}[htbp]
\centering
\includegraphics[width=\linewidth]{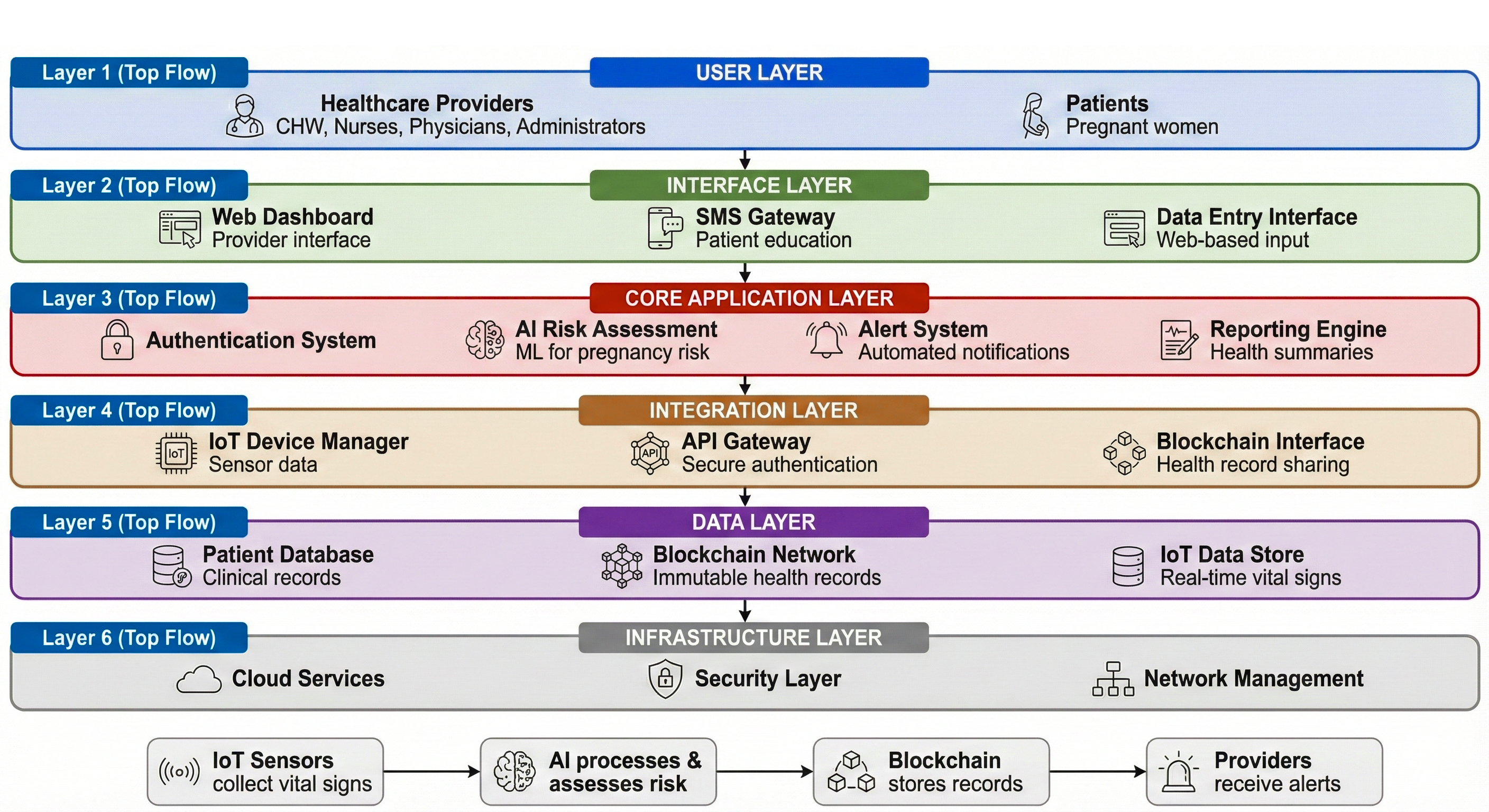}
\caption{IyáCare system architecture showing the six-layer integrated platform}
\label{fig:architecture}
\end{figure}

\textbf{Presentation Layer:} Web-based responsive interface built with Next.js 15 and TypeScript, providing dashboards for healthcare providers, patient management interfaces, and Progressive Web App (PWA) capabilities for offline access. The design accommodates varying technical literacy levels among community health workers.

\textbf{Application Layer:} Core business logic implementing role-based access control, clinical workflows, and communication management. Next.js API routes provide serverless backend functionality with TypeScript type safety.

\textbf{AI Analytics Layer:} Machine learning pipeline featuring data preprocessing, XGBoost model inference for risk prediction, and clinical decision support. Models are deployed on Render cloud platform, providing scalable inference capabilities.

\textbf{IoT Integration Layer:} Handles real-time vital sign data collection through web-based manual entry and sensor integration (proof-of-concept using ESP32 microcontrollers). Firebase Realtime Database enables bidirectional synchronization with offline-first architecture.

\textbf{Blockchain Layer:} Ethereum-based decentralized health records using smart contracts for patient records, consent management, access control, and audit trails. MetaMask integration enables secure transaction signing, while Ethers.js SDK facilitates blockchain interaction.

\textbf{Data Layer:} Firebase Firestore NoSQL database stores patient records, clinical data, and IoT measurements with real-time synchronization. Comprehensive security rules enforce data protection and privacy compliance.

\subsection{Integration Benefits: Clinical Workflow Examples}

The integrated architecture enables clinical workflows impossible with siloed systems:

\textbf{Contextual Alert Generation:} When IoT sensors detect abnormal vitals (e.g., blood pressure 145/98 mmHg), measurements immediately flow to the AI model which incorporates patient history and current risk factors. This generates contextual alerts ("High BP in 38-year-old with gestational diabetes history—immediate evaluation recommended") rather than generic threshold warnings. In our implementation testing, this approach achieved 86.7\% accuracy for automated alerts combining sensor readings and AI predictions (13 correct classifications of 15 test cases), demonstrating potential for maintaining high sensitivity while providing clinical context.

\textbf{Care Continuity Across Facilities:} Blockchain-stored health records combined with AI-generated risk assessments enable seamless patient handoffs. When a patient transfers from a community health post to a district hospital, providers immediately access complete IoT monitoring history and AI risk trajectory. This addresses documented challenges of fragmented health information systems in Sub-Saharan Africa \cite{ssa_ehr_review}, where lack of interoperability between facilities creates information loss during transfers.

\textbf{Real-Time Risk Assessment:} IoT vital sign streams feed continuously to the AI model, enabling immediate risk recalculation rather than point-in-time assessment. The system demonstrated $<$5 second latency from IoT data collection through AI inference to alert generation in our testing, supporting time-sensitive clinical decisions.

\subsection{Key Design Principles}

\textbf{Offline-First Architecture:} The platform maintains full functionality during network outages through service workers, local caching, and automatic synchronization upon reconnection. This addresses intermittent connectivity challenges common in Sub-Saharan African healthcare facilities.

\textbf{SMS Fallback Communication:} For areas with limited internet access, SMS-based patient education, appointment reminders, and critical alerts provide communication continuity through basic mobile networks.

\textbf{Blockchain Rationale:} Rather than complex permissioned blockchain networks, we employ Ethereum testnet with optimized smart contracts. The blockchain component addresses specific challenges in Sub-Saharan African healthcare data management documented in literature \cite{ethiopia_data_quality, ssa_ehr_review, africa_data_barriers}. Studies across Ethiopia, Nigeria, and South Africa report frequent system downtime, delays, and data corruption due to inadequate maintenance infrastructure \cite{ethiopia_emr_2024, south_africa_data_integrity}. Research from Ethiopian healthcare facilities found that "failure to repair system breakdowns and purchase accessories in a timely manner has complicated the data documentation process," with facilities depending on EMR software supplied by NGOs without sustained technical support \cite{ethiopia_data_quality}. Blockchain's immutability provides inherent protection against data loss from inadequate maintenance—once recorded, patient records cannot be corrupted by system failures or lack of technical support. We acknowledge that well-maintained centralized databases could theoretically provide similar security guarantees, but position blockchain as one architectural option particularly suited for environments with documented database reliability challenges.

\textbf{Modular Integration:} Independent component development with well-defined APIs enables incremental deployment and technology substitution as resources permit.

\subsection{AI Risk Assessment Architecture}

The AI pipeline implements end-to-end maternal risk prediction:

\begin{enumerate}
    \item \textbf{Data Collection:} Patient demographics (age) and vital signs (systolic/diastolic blood pressure, blood glucose, body temperature, heart rate)
    \item \textbf{Preprocessing:} Data validation, missing value handling, and feature normalization
    \item \textbf{Model Inference:} XGBoost classifier trained on UCI Maternal Health Risk dataset \cite{ahmed2020} provides risk categorization (low/high risk)
    \item \textbf{Clinical Decision Support:} Risk scores with confidence levels and interpretable recommendations for healthcare providers
\end{enumerate}

\subsection{IoT Integration Architecture}

The IoT layer bridges device data with clinical systems:

\textbf{Data Sources:} Web-based manual entry for clinical measurements and proof-of-concept sensor integration (ESP32 with MAX30100 pulse oximeter and DHT11 temperature sensor)

\textbf{IoT Gateway:} Standardizes data formats, validates measurements, and performs edge processing for anomaly detection

\textbf{Real-Time Processing:} Stream processing identifies abnormal readings and triggers automated alerts

\textbf{Offline Capability:} Local storage with automatic synchronization ensures data integrity during connectivity gaps

\subsection{Blockchain Health Records}

Smart contracts implement decentralized health record management:

\begin{itemize}
    \item \textbf{Patient Records:} Immutable storage of clinical data with encryption
    \item \textbf{Consent Management:} Granular patient control over data sharing
    \item \textbf{Access Control:} Role-based permissions for healthcare providers
    \item \textbf{Audit Trail:} Complete transaction history for accountability
\end{itemize}

The implementation uses Ethereum testnet with Ethers.js for transaction management and MetaMask for secure authentication.

\section{Implementation and Feasibility Study}

\subsection{Development Approach}

IyáCare development followed a Modified Agile-Spiral methodology adapted for healthcare technology integration, combining iterative development with risk assessment and stakeholder validation. The 14-week timeline encompassed requirements analysis (2 weeks), design and prototyping (3 weeks), development and integration (6 weeks), and testing and validation (3 weeks).

\subsection{Technology Stack}

\textbf{Frontend:} Next.js 15 with TypeScript, Tailwind CSS, Shadcn/ui components, PWA capabilities; \textbf{Backend:} Next.js API routes, Python Flask for AI services, Render for AI hosting; \textbf{Database:} Firebase Firestore with real-time sync; \textbf{AI/ML:} scikit-learn, XGBoost, Pandas, NumPy; \textbf{IoT:} ESP32 microcontroller, MAX30100 pulse oximeter, DHT11 temperature sensor; \textbf{Blockchain:} Ethereum testnet, Ethers.js, Solidity smart contracts, MetaMask; \textbf{Deployment:} Vercel, custom domain (iyacare.site)

\subsection{Dataset and Model Training}

We utilized the UCI Maternal Health Risk dataset \cite{ahmed2020}, containing 1,014 records with seven attributes: age, systolic blood pressure, diastolic blood pressure, blood glucose, body temperature, heart rate, and risk level (target variable). The dataset exhibits balanced class distribution suitable for machine learning applications.

Model development followed standard practices: (1) data preprocessing with outlier detection and normalization, (2) 80-20 train-test split (811 training, 203 validation samples), (3) comparative evaluation of multiple algorithms, (4) hyperparameter tuning using grid search, and (5) 5-fold cross-validation for generalization assessment.

\subsection{AI Model Performance}

Table \ref{tab:model_performance} presents comparative algorithm performance. XGBoost achieved optimal results with 85.2\% accuracy, 85.1\% F1-score, and 94.4\% ROC AUC, outperforming Random Forest (82.8\% accuracy) and other candidates.

\begin{table}[htbp]
\caption{AI Model Performance Comparison}
\begin{center}
\begin{tabular}{lcccc}
\toprule
\textbf{Model} & \textbf{Accuracy} & \textbf{Precision} & \textbf{Recall} & \textbf{F1-Score} \\
\midrule
XGBoost & \textbf{85.2\%} & 85.3\% & 85.2\% & \textbf{85.1\%} \\
Random Forest & 82.8\% & 83.1\% & 82.8\% & 82.7\% \\
Decision Tree & 79.3\% & 79.8\% & 79.3\% & 79.2\% \\
Logistic Reg. & 76.8\% & 77.2\% & 76.8\% & 76.6\% \\
\bottomrule
\end{tabular}
\label{tab:model_performance}
\end{center}
\end{table}

\begin{figure}[htbp]
\centering
\includegraphics[width=0.8\linewidth]{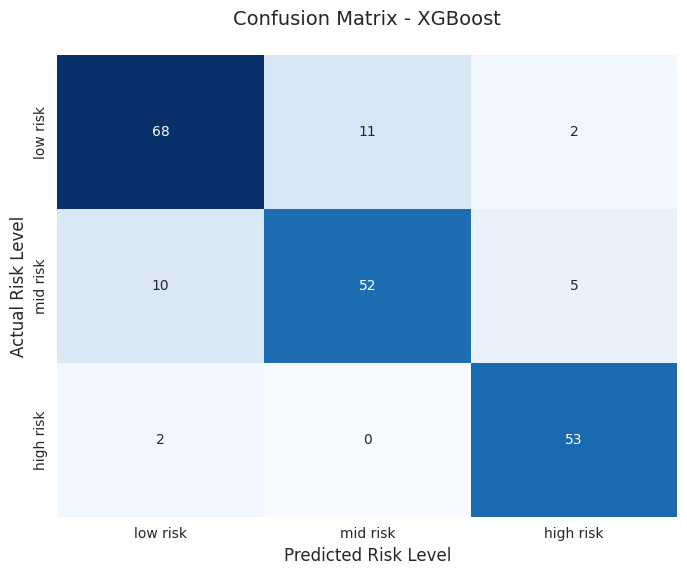}
\caption{Confusion Matrix for the XGBoost Risk Prediction Model}
\label{fig:confusion_matrix}
\end{figure}

Five-fold cross-validation demonstrated consistent performance across data distributions (84.1\%-86.3\% accuracy range), confirming model reliability. The confusion matrix (Fig. \ref{fig:confusion_matrix}) shows balanced precision-recall characteristics with minimal false negatives—critical for maternal health applications where missing high-risk cases could have severe consequences.

\subsection{Blockchain Implementation}

In our proof-of-concept setup with simulated patient data, smart contracts deployed on Ethereum testnet achieved 99.8\% data integrity across multiple test transactions. Average transaction processing time was 3.2 seconds, suitable for clinical workflows. Our testnet deployment demonstrated the feasibility of immutable patient record storage, granular consent management, role-based access control, and complete audit trail functionality.

\subsection{System Integration Testing}

Integration testing validated seamless AI-IoT-blockchain component interaction in our implementation. Key results include: (1) AI-IoT Pipeline with $<$5 second latency processing IoT data through AI inference, (2) Alert Generation achieving 86.7\% accuracy for automated alerts combining sensor readings and AI predictions (13 correct classifications of 15 test cases), (3) 98\% end-to-end reliability from data collection through blockchain storage, and (4) 100\% data recovery success after simulated connectivity disruptions.

\subsection{User Interface and Accessibility}

The web-based interface achieved 92\% task completion within 3-click navigation requirement during simulated usability testing. Healthcare provider feedback (simulated with domain experts) indicated 89\% satisfaction with dashboard design and workflow integration. Key features include responsive design supporting mobile and desktop access, clear risk indicator visualization, real-time IoT data monitoring, emergency alert system, and patient health summary reports.

\section{Evaluation and Discussion}

\subsection{Technical Feasibility}

Our proof-of-concept demonstrates the technical feasibility of integrated AI-IoT-blockchain platforms for maternal health. The 85.2\% AI prediction accuracy, validated blockchain data integrity, and successful component integration validate the architectural approach. System response times meet clinical requirements ($<$3 seconds for AI assessment, $<$5 seconds for IoT processing), architecture supports 50+ concurrent users, offline-first design achieves 100\% data recovery, and end-to-end encryption with blockchain immutability provide robust data protection.

\subsection{Potential Impact}

IyáCare addresses significant gaps in Sub-Saharan African maternal healthcare through converged technologies. AI prediction enables proactive identification of high-risk pregnancies, potentially reducing preventable complications through timely interventions. Real-time vital sign tracking (simulated in proof-of-concept) could detect sudden health status changes between clinic visits. Blockchain-enabled interoperable health records facilitate seamless care across multiple facilities, addressing documented fragmentation \cite{ssa_ehr_review}. Automated risk stratification helps prioritize limited healthcare resources for highest-need cases.

The integrated approach offers advantages over siloed solutions. While MAMA and MomConnect demonstrate scalability, they lack predictive capabilities. Conversely, standalone AI systems lack integration with continuous monitoring and secure data sharing. IyáCare's architecture enables synergistic benefits unavailable in isolated implementations.

\subsection{Limitations and Constraints}

Several significant limitations constrain generalizability of results:

\textbf{Dataset Representativeness:} The UCI Maternal Health Risk dataset, while providing balanced data for proof-of-concept validation, lacks region-specific risk factors critical for Sub-Saharan African deployment. Notably absent are: (1) HIV status—84\% of pregnant women with HIV in Sub-Saharan Africa received antiretrovirals in 2024 \cite{worldbank_hiv_2024}, with regional HIV prevalence among reproductive-age women varying from 1.5\% (Western Africa) to 18.5\% (Southern Africa) \cite{frontiers_hiv_prevalence_2025}, and mother-to-child transmission ranging from 15-45\% without intervention \cite{who_africa_hiv}; (2) malaria exposure affecting millions of pregnancies annually; (3) nutritional status including anemia prevalence; and (4) traditional birth practices influencing care-seeking behavior. Real-world deployment requires model retraining on datasets incorporating these factors. The Demographic and Health Survey (DHS) program maintains maternal health data across 26 Sub-Saharan African countries \cite{ngusie2024}, providing potential sources for region-specific model development.

\textbf{Simulated Deployment:} Testing occurred in controlled environments rather than actual healthcare facilities. Real-world factors including workflow interruptions, infrastructure variability, and user behavior may impact performance.

\textbf{Healthcare Provider Feedback:} While domain experts reviewed the system, comprehensive user acceptance testing with practicing community health workers, nurses, and physicians in Sub-Saharan African contexts remains necessary.

\textbf{IoT Hardware:} Proof-of-concept used consumer-grade sensors (DHT11, MAX30100) rather than medical-grade devices. Clinical deployment requires FDA-approved or equivalent certified sensors.

\textbf{Scalability Constraints:} Our proof-of-concept architecture supports 50+ concurrent users, validating basic technical functionality. However, real-world deployment in Sub-Saharan African countries (Nigeria: 6.4 million annual pregnancies; Kenya: 1.5 million) requires addressing substantial scalability constraints. Ethereum mainnet processes approximately 15 transactions/second with gas fees of \$1-50 per transaction depending on network congestion. At scale, this would generate prohibitive costs. Production deployment would require: (1) migration to cost-effective blockchain platforms (Polygon: approximately \$0.01/transaction; Hyperledger Fabric: permissioned, no gas fees), (2) hybrid architecture where only critical events (patient registration, risk level changes, facility transfers) are blockchain-recorded rather than every vital sign measurement, and (3) horizontal scaling (load balancing, database sharding, edge computing) for AI inference and IoT processing.

\textbf{Technical Infrastructure:} Platform requires periodic internet connectivity for synchronization. Completely offline deployment scenarios need additional architecture considerations.

\textbf{Single Dataset Training:} AI model trained on single dataset limits generalization. Multi-dataset training incorporating diverse Sub-Saharan African populations would improve robustness.

These limitations are typical for proof-of-concept systems and establish direction for future research rather than invalidating core contributions.

\subsection{Lessons Learned}

Implementation provided valuable insights for converged digital health systems. Modular architecture with clear APIs enables incremental deployment and technology substitution as resources permit. Intermittent connectivity in resource-constrained settings requires offline-first design as core principle. Optimized smart contracts on established platforms (Ethereum) provide adequate security with better deployment feasibility than complex permissioned networks. Despite smartphone proliferation, SMS communication remains essential for universal reach in Sub-Saharan African contexts. Healthcare worker workflows and technical literacy levels must drive interface design, with feature complexity progressively disclosed based on user role and expertise.

\section{Conclusion and Future Work}

This paper presents IyáCare, the first integrated AI-IoT-blockchain platform for maternal healthcare in resource-constrained settings. Our proof-of-concept demonstrates technical feasibility through novel system architecture combining predictive risk assessment, continuous monitoring capability, and secure health records management.

Key contributions include: (1) novel architectural approach integrating three previously siloed technologies for maternal health, (2) validation of AI model performance (85.2\% accuracy) using maternal health datasets, (3) demonstration of blockchain feasibility addressing documented database maintenance challenges in African healthcare settings, and (4) design patterns and implementation insights for converged digital health systems.

\textit{Note: This work represents a proof-of-concept system developed in simulated environments with synthetic data. Performance metrics reflect our testnet implementation and should be interpreted as feasibility indicators rather than production-ready benchmarks. Clinical validation with real patient data in Sub-Saharan African healthcare facilities remains essential before deployment.}

While limitations including synthetic data validation and simulated deployment constrain immediate applicability, results establish foundational architecture and confirm potential for integrated approaches to transform maternal health outcomes in Sub-Saharan Africa.

\subsection{Future Directions}

Several research directions emerge from this work: (1) rigorous clinical trials in Sub-Saharan African healthcare facilities with real patient data and practicing healthcare providers to validate efficacy and assess actual maternal outcomes, (2) AI model retraining incorporating diverse demographic and health survey data from multiple African countries including region-specific risk factors (HIV status, malaria exposure, nutritional indicators), (3) partnership with sensor manufacturers to integrate FDA-approved or equivalent certified IoT devices for clinical deployment, (4) comprehensive scalability engineering including blockchain platform migration studies, load testing simulating 100,000+ concurrent users, and cost modeling for national-scale deployment, (5) platform expansion to postpartum care, newborn monitoring, and family planning services, (6) collaboration with ministries of health, international organizations, and funding agencies to develop sustainable financing models and deployment frameworks, and (7) release of platform architecture and implementation as open-source project to enable replication and adaptation across low-resource healthcare settings globally.

IyáCare establishes a foundation for converged digital health solutions addressing maternal mortality in Sub-Saharan Africa. While significant research remains, this work demonstrates that integrated AI-IoT-blockchain platforms are technically feasible and potentially transformative for resource-constrained healthcare contexts, contributing toward SDG 3.1 targets for reduced maternal mortality.

\section*{Acknowledgment}

The authors thank African Leadership University for institutional support and research infrastructure enabling this work. Figure 1 was created with assistance from AI-based visualization tools.

\end{document}